\documentclass[conference]{article}  

\usepackage{graphics} 
\usepackage{epsfig} 
\usepackage{color}
\usepackage{booktabs}
\usepackage{mathptmx} 
\usepackage{times} 
\usepackage{amsmath} 
\usepackage{amssymb}  
\usepackage{dsfont}
\usepackage{xurl}
\usepackage{a4wide}
\usepackage{bm} 

\usepackage{algorithm}
\usepackage{algpseudocode}

\newtheorem{assumption}{Assumption}

\newtheorem{definition}{Definition}

\newcommand{\red}[1]{\textcolor{red}{#1}}

\title{\LARGE \bf
Data-driven port-Hamiltonian structured identification \\ for non-strictly passive systems
}

\author{Charles Poussot-Vassal$^{1,\dagger}$, Denis Matignon$^{2}$, Ghilslain Haine$^{2}$ and Pierre Vuillemin$^{1}$
\thanks{*This work has been supported by the AID (Agence de l’Innovation de Défense) from the French Ministry of the Armed Forces (Ministère des Armées).
}
\thanks{$^{1}$Charles and Pierre are with ONERA -- The French Aerospace Lab, 2 avenue Edouard Belin, Toulouse 31400, France.}
\thanks{$^{2}$Denis and Ghislain are with ISAE-SUPAERO, Université de Toulouse, 10, avenue Edouard Belin, Toulouse 31400, France.}
\thanks{$^\dagger$ Corresponding author: {\tt\footnotesize charles.poussot-vassal@onera.fr}}%
}

\newcommand{\grad}{\mathbf{grad}}

 \newcommand{\aq}{\boldsymbol{\alpha}_q}
\newcommand{\eq}{\mathbf{e}_q}
\newcommand{\phiq}{\boldsymbol{\varphi}_q}

\providecommand{\eg}{\emph{e.g. }}
\providecommand{\ie}{\emph{i.e. }}
\providecommand{\System}{\boldsymbol \Sigma}

\providecommand{\Systemr}{\boldsymbol{\hat \Sigma}}
\providecommand{\Er}{\hat E}
\providecommand{\Ar}{\hat A}
\providecommand{\Br}{\hat B}
\providecommand{\Cr}{\hat C}

\providecommand{\SystemPh}{\boldsymbol{\Sigma}_{\text{pH}}}

\providecommand{\LL}{\mathbb L}
\providecommand{\sLL}{\mathbb M}
\providecommand{\diag}{\textbf{diag}}
\providecommand{\rank}{\textbf{rank}}
\providecommand{\Htran}{\mathbf{H}}

\providecommand{\Htranr}{\mathbf{\hat H}}
\providecommand{\Hreal}{\boldsymbol \Sigma}
\providecommand{\Hrealr}{\hat{\Hreal}}
\providecommand{\matrixtwo}[4]{ \left[\begin{array}{cc} #1 & #2 \\ #3 & #4 \end{array}\right] } %

\providecommand{\denis}[1]{\red{#1}}

\begin{document}

\maketitle
\thispagestyle{empty}
\pagestyle{empty}

\begin{abstract}
In this work, we detail a procedure to construct a reduced order model on the basis of frequency-domain data, that preserves the non-strictly passive property and the port-Hamiltonian structure. The proposed scheme is based on Benner et al. contribution \cite{Benner:2020}, which has been adapted (i) to handle non-strictly passive model, and (ii) to handle numerical issues observed when applying the Loewner framework on complex configurations. We validate the proposed scheme on a very complex two-dimensional wave equation, for which the discretized version preserves the port-Hamiltoninan form. 
\end{abstract}


\section{Introduction}
\subsection{Motivation and foreword}


The present work  is motivated by an efficient numerical representation of the wave equation on a 2D domain $\Omega$, with  actuators and sensors that are \emph{collocated} at the boundary $\partial\Omega$; the PDE model is first described as a distributed port-Hamiltonian system (pHs) (see  the pioneering work \cite{van2002hamiltonian},  and \cite{rashad2020twenty} for a recent  overview), second discretized in a structure-preserving manner thanks to the Partitioned Finite Element Method (PFEM) \cite{CarMatLef21}. Although in 1D with physical parameters 
which are uniform in space, 
the input-output transfer function is easy to compute,  
the task becomes more difficult with varying parameters. In generic geometric 2D domain with heterogeneous and anisotropic parameters, it is almost impossible. However, a high-fidelity full order model (FOM), taking all these important properties into account can be computed at the discrete level \cite{SerMatHaiGSI19}.

This model results in a linear pH one, embedding a very large number of internal state variables, a quite large number of inputs and outputs. 
Such a high dimension is a limiting factor for simulation, optimisation, analysis and control. 
Computing simplified, easy to use dynamical models is one purpose of the model approximation and reduction discipline. 
The goal is to approximate the original system with a smaller and simpler system with the \emph{same structure} and similar response characteristics as the original, the low-complexity model, also called a reduced order model (ROM).  

The Loewner framework (LF) employed in this work is a \emph{data-driven} model identification and reduction technique that was originally introduced in \cite{Mayo:2007}. Using only frequency-domain measured data, the LF constructs surrogate models directly and with low computational effort. Its extension to pH model is proposed in \cite{Benner:2020} and \cite{AntoulasSurvey:2016}. We refer the reader to \cite{GoseaHNA:2022} for an overview of Loewner identification and reduction methods. 
Recently \cite{GhattasWillcoxActa2021} gave an overview of physics-based reduction methods, and \cite{CherifiBrugnoliLHMNLC21} presents a successful attempts to apply data-driven techniques to identification of pHs on a 1D example. 

\subsection{Notations and preliminaries}
The set of real and complex numbers of dimension $n$ are denoted, respectively by $\mathbb R^n$ and $\mathbb{C}^n$. The complex variable $\imath=\sqrt{-1}$. The notation $\mathbb X^n_\Lambda:\{x\in \mathbb X^n \setminus \Lambda\}$, where $\Lambda$ denotes a finite number set (typically singularities) in $\mathbb X^n$ ($\mathbb X^n=\{\mathbb R^n,\mathbb C^n\}$). The set of stable rational functions with bounded $\infty$-norm  along $\imath \mathbb R$, is denoted $\mathcal{RH}_\infty$. Similarly,  $\mathcal{RL}_\infty$ denotes the same set but for both stable and unstable functions. Identity and null matrices of dimension $p$ read $I_p$ and $0_p$. The Laplace variable is denoted $s\in\mathbb C$. Here we consider the following multi-input multi-output (MIMO) linear time invariant (LTI) continuous-time dynamical systems realisations (with $x(0)=0$):
\begin{subequations}
\begin{equation}
    E\dot x(t)=Ax(t)+Bu(t) ,\quad
    y(t)=Cx(t),
\label{eq:EABC}
\end{equation}
\begin{equation}
    \dot x(t)=Ax(t)+Bu(t)  ,\quad
    y(t)=Cx(t)+Dx(t),
\label{eq:ABCD}
\end{equation}
\begin{equation}
\left\{
\begin{array}{rcl}
    M\dot x(t)&=&(J-R)Qx(t)+(G-P)u(t) \\
    y(t)&=&(G+P)^\top Qx(t)+(N+S)u(t)
\end{array}
\right.,
\label{eq:pH}
\end{equation}
\end{subequations}
where $x(t)\in\mathbb R^n$ and $u(t),y(t)\in\mathbb R^m$ are vector-valued functions denoting the internal variables, input and output of the system. In the standard descriptor \eqref{eq:EABC} and non-descriptor \eqref{eq:ABCD} forms, we consider constant matrices $E,A\in\mathbb R^{n\times n}$, $B,C^\top \in\mathbb R^{n\times m}$ and $D\in\mathbb R^{m\times m}$. When considering the port-Hamiltonian form \eqref{eq:pH}, $M,J,R,Q\in\mathbb R^{n\times n}$, $G,P\in\mathbb R^{n\times m}$ and  $N,S\in\mathbb R^{m\times m}$. For brevity, \eqref{eq:EABC} and \eqref{eq:ABCD} are denoted $\System:=(E,A,B,C,0_m)$ and $\System:=(I_n,A,B,C,D)$ respectively. The pH form \eqref{eq:pH} is shortly denoted $\SystemPh:=(M,Q,J,R,G,P,N,S)$. By introducing the co-energy variable $e(t)=Qx(t)$, \eqref{eq:pH} boils down to $M\dot x(t)=(J-R)e(t)+(G-P)u(t)$ and $y(t)=(G+P)^\top e(t)+(N+S)u(t)$. The latter is the so-called \emph{co-energy pH form} and is of specific meaning in the computation of the Hamiltonian. 
In each case, we define the associated transfer functions as $\Htran: \mathbb C_\Lambda \mapsto \mathbb C^{m\times m}$, where $\Htran(s) = C(sE-A)^{-1}B$ for \eqref{eq:EABC}, $\Htran(s) = C(sI-A)^{-1}B+D$ for \eqref{eq:ABCD} and $\Htran(s) = (G+P)^\top Q(sM-(J-R)Q)^{-1}(G-P)+(N+S)$ for \eqref{eq:pH}\footnote{Here $\Lambda$ denotes the singularities being the eigenvalues of $(A,E)$ pencil in \eqref{eq:EABC}, of $A$ in \eqref{eq:ABCD} and of $((J-R)Q,M)$ in \eqref{eq:pH}.}. On the basis of $\Htran$, let us denote the spectral density as 
$\boldsymbol \Phi_\Htran(s) := \Htran(s) + \Htran^\top (-s)$ and let us remind the following definitions, necessary to characterise a pH system.
\begin{definition}[Positive realness]
For all $\omega \in \mathbb R$, the rational transfer $\Htran(s)$ is called strictly positive-real if $\boldsymbol \Phi_\Htran(\imath\omega) \succ 0$ and positive-real if $\boldsymbol \Phi_\Htran(\imath\omega) \succeq 0$.
\end{definition}

\begin{definition}[Stability]
The rational transfer function $\Htran(s)$ is called asymptotically stable if its singularities $\Lambda$ are in the open left-half plane, and called stable it its singularities $\Lambda$ are in the closed left-half plane with any pole occurring on $\imath\mathbb R$ being not repeated.
\end{definition}

\begin{definition}[Passivity]
The rational transfer function $\Htran(s)$ is called strictly passive if it is strictly positive real and asymptotically stable, and stable if positive real and stable.
\end{definition}

\subsection{Contribution statement and paper organisation} 

This note is grounded on the LF 
extended by \cite{Benner:2020} to identify pH-ROM as defined in \eqref{eq:pH}. The contributions are twofold: \emph{(i)} first, to propose both a methodological and numerical adjustment from \cite{Benner:2020}'s algorithm to identify pH models where the original system is passive but not strictly\footnote{One should also point that this is also treated in \cite{Breiten:2022} through a model-based approach via what authors call the \emph{regular} and \emph{singular} cases.} 
\emph{(ii)} and second, to apply the proposed process to a highly dimensional ($n\gg10^4$) and MIMO system, embedding a rich dynamic: the 2D wave equation, which complexity is far higher than standard benchmarks. 
The paper is organised as follows: in \S \ref{sec:rom}, the proposed data-driven pH-ROM construction method for non-strictly passive systems is presented, using an adaptation of the data-driven LF of \cite{Benner:2020}. The approach is illustrated and validated through a numerical example resulting from a complex 2D wave equation in \S \ref{sec:num}. Conclusions and perspectives are drawn in \S \ref{sec:conclusion}.


\section{Port-Hamiltonian identification in the Loewner framework}
\label{sec:rom}

We are interested in identifying a MIMO ROM preserving the pH structure of $\SystemPh$, using a data-driven framework. To do so, we follow the approach of \cite{Benner:2020} which extends the LF originally presented in \cite{Mayo:2007}. The latter is first reminded in \S \ref{sub:loe}, while the former is presented in \S\ref{sub:loe_passive}. The proposed algorithm allowing to cope with non-strictly passive systems, is detailed in \S \ref{sub:loe_passive2}.


\subsection{Loewner framework preliminaries}
\label{sub:loe}

The LF offers tools for the reduction, approximation and identification of dynamical systems based on frequency-domain data. Let us denote as the \emph{right and left data} the following sets  (where $j=1, \dots, k$ and $i =1, \dots,q$):
\begin{equation}
    \{\lambda_j, \mathbf{r}_j, \mathbf{w}_j\} 
    \text{ and }
    \{\mu_i, \mathbf{l}_i^\top , \mathbf{v}_i^\top \}, 
    \label{eq:data}
\end{equation}
where $\lambda_j\in \mathbb C$ and $\mu_i\in \mathbb C$ are the right  and left interpolation points. Then,  $\mathbf{r}_j\in\mathbb C^{m\times 1}$ and $\mathbf{l}_i^\top \in\mathbb C^{1\times m}$ are the right and left tangential directions. Both points and directions lead to the right $\Htran(\lambda_j)\mathbf r_j=\mathbf{w}_j\in\mathbb C^{m\times 1}$ and left  $\mathbf l_i^\top \Htran(\mu_i) = \mathbf{v}_i^\top \in\mathbb C^{1\times m}$ tangential responses. $\Htran(s_k)$ is the evaluation of the high dimensional pH-FOM at point $s_k\in\mathbb C$. Based on \eqref{eq:data}, the LF seeks for $\Hrealr: (\Er,\Ar,\Br,\Cr,0_m)$, whose transfer function  $\Htranr(s)$ satisfies tangential interpolatory conditions $\Htranr(\lambda_j)\mathbf r_j = \mathbf{w}_j$ and $\mathbf l_i^\top \Htranr(\mu_i) = \mathbf{v}_i^\top $. By using the matrix formulation, the right  data read
\begin{equation}
    \left\{
    \begin{array}{rcl}
    \Lambda &=&~\diag~[\lambda_1, \dots, \lambda_k]\in \mathbb{C}^{k\times k}, \\ 
    \mathbf{R} &=& \begin{bmatrix}
    \mathbf{r}_1 & \mathbf{r}_2 & \dots & \mathbf{r}_k \end{bmatrix} \in \mathbb{C}^{m \times k} \\
    \mathbf{W} &=& \begin{bmatrix}
    \mathbf{w}_1 & \mathbf{w}_2 & \dots & \mathbf{w}_k \end{bmatrix} \in \mathbb{C}^{m \times k}
    \end{array}
    \right. ,
\end{equation}
and the left data read
\begin{equation}
    \left\{
    \begin{array}{rcl}
    \mathbf{M} &=&~\diag~[\mu_1, \dots, \mu_q] \in \mathbb{C}^{q\times q} \\ 
    \mathbf{L}^\top &=& \begin{bmatrix}
    \mathbf{l}_1 & \mathbf{l}_2 & \dots & \mathbf{l}_q \end{bmatrix} \in \mathbb{C}^{m \times q} \\ \mathbf{V}^\top  &=& \begin{bmatrix}
    \mathbf{v}_1 & \mathbf{v}_2 & \dots & \mathbf{v}_q \end{bmatrix} \in \mathbb{C}^{m \times q} 
    \end{array}
    \right. .
\end{equation}
Then, by defining the $i,j$-th entry of the Loewner and shifted Loewner matrices as
\begin{equation}
    (\LL)_{ij} =  \dfrac{\mathbf{v}_i^\top \mathbf r_j-\mathbf{l}_i^\top \mathbf{w}_j}{\mu_i -\lambda_j}
    \text{ and }
    (\sLL)_{ij} = \dfrac{\mu_i\mathbf{v}_i^\top \mathbf r_j-\mathbf{l}_i^\top \mathbf{w}_j\lambda_j}{\mu_i -\lambda_j},
    \label{eq:loewner}
\end{equation}
the resulting system realization $\Systemr:=(\Er,\Ar,\Br,\Cr,0_m)=(-\LL,-\sLL,\mathbf V,\mathbf W,0_m)$ which transfer function $\Htranr(s)=\mathbf W(\sLL-s\LL)^{-1}\mathbf V$ (tangentially) interpolates the \emph{data}. It follows that Loewner matrices satisfy the Sylvester equations $\mathbf{M}\mathbb{L}- \mathbb{L}\boldsymbol{\Lambda}  = \mathbf{V}\mathbf{R} - \mathbf{L}\mathbf{W}$ and 
$\mathbf{M}\sLL  -\sLL\boldsymbol{\Lambda} = \mathbf{M}\mathbf{V}\mathbf{R} - \mathbf{L}\mathbf{W}\boldsymbol{\Lambda}$. If data have been generated by a linear rational model, $\forall \xi \in\mathbb C\setminus\Lambda$, the rational order $r = \rank (\xi \LL- \sLL) = \rank ([\LL,\sLL]) =  \rank ([\LL^H,\sLL^H]^H)$ recovers the the minimal realisation of the generating system, as well as its McMillan degree $\nu = \rank(\LL)$. 
These features make this approach central in the realisation theory (see \cite{AntoulasSurvey:2016,GoseaHNA:2022} for a recent overviews).

\subsection{Loewner framework with strict passivity}
\label{sub:loe_passive}

\subsubsection{General ideas and assumptions}

Applying the LF to \emph{data} generated by a passive system $\Htran$ do not necessarily lead to a passive transfer $\Htranr$. This is solved in \cite{Benner:2020} by through specific \emph{right and left data} selection. This result is recalled here together with the main (limiting) assumptions.

\begin{assumption}[Strictly passive]
\label{ass:strict_passivity}
In \cite{Benner:2020}, authors assume that the system generating the data to be strictly passive, implying proper transfer matrix $\Htran$ where singularities are not on the imaginary axis or at infinity. 
\end{assumption}

\begin{assumption}[Stability]
\label{ass:stability}
In \cite{Benner:2020}, authors assume that model $\Htranr$ (realisation $\Systemr$) obtained after a first identification in the LF leads to a stable pencil $(\sLL,\LL)=(\Ar,\Er)$, \ie  $\Lambda\in\mathbb C_-$.
\end{assumption}

\subsubsection{Procedure as given in \cite{Benner:2020}}

First, let $\Htranr$ be identified by the LF, on the basis of a real and strictly passive transfer function $\Htran$, where the $D$-term is removed to avoid rank deflecting $\Er=\LL$ matrix. It results in $\Htranr$ where McMillan degree $\nu$ is equal to the (minimal) realisation order $r$, since no polynomial term appear (because of the strict passivity and $D$-term removal). Note that $r$ may be automatically selected by the rank revealing factorisation of the LF or be chosen smaller. This identified model realisation $\Systemr$ is now used to estimate the associated spectral zeros and directions pairs, denoted $(\xi_j,\mathbf x_j)$ such that
$\boldsymbol \Phi_\Htranr(\xi_j)\mathbf x_j = 0$. This pair is computed by solving the following
low order generalized eigenvalue problem (see \cite{Willems:1972,Benner:2020}):
\begin{equation}
    \begin{bmatrix}
    0 & \Ar & \Br \\ \Ar^\top  & 0 & \Cr^\top  \\ \Br^\top  & \Cr & D+D^\top 
    \end{bmatrix}
    \begin{bmatrix}
    p_j\\ q_j\\ \mathbf x_j
    \end{bmatrix}
    = \xi_j
    \begin{bmatrix}
    0 & \Er & 0\\ -\Er^\top  & 0 & 0\\ 0 &0 &0 
    \end{bmatrix}
    \begin{bmatrix}
    p_j\\ q_j\\ \mathbf x_j
    \end{bmatrix}.
    \label{eq:SZ}
\end{equation}
According to Assumptions \ref{ass:strict_passivity} and \ref{ass:stability}, this eigen-problem has $r$ zeros in the open right half-plane, $r$ zeros in the open left half-plane and has no zeros on the imaginary axis. By selecting the \emph{right  and left strictly passive data} data as ($i,j=1,\dots, r=k=q$,  $\lambda_j\leftarrow \xi_j$ and $\mathbf{r}_j\leftarrow \mathbf{x}_i$), 
\begin{equation}
    \{\lambda_j, \mathbf{r}_j, \mathbf{w}_j\} 
    \text{ and }
    \{-\overline \lambda_i, \mathbf{r}_i^H, -\mathbf{w}_i^H\}, 
    \label{eq:data_strict_passive}
\end{equation}
one gets, $\mathbf M=-\boldsymbol \Lambda^H$, $\mathbf L=\mathbf R$ and $\mathbf V=-\mathbf W^H$. Therefore, by construction, one obtains an Hermitian $\LL\in\mathbb C^{r\times r}$ and a skew symmetric $\sLL\in\mathbb C^{r\times r}$ matrix \eqref{eq:loewner}. By setting, $\Htran(\infty)=D$ (which may be estimated by sampling in very high frequency), one recovers an $m\times m$ real transfer function $\Htranr$\footnote{Real matrices are obtained if data are sampled with complex conjugate frequencies and by applying a unitary projection  \cite{AntoulasSurvey:2016,GoseaHNA:2022}.}. As $\LL\succ0$, one may apply the Cholesky decomposition $\LL=T^\top  T$. Then the \emph{normalized pH model} is obtained as $\System_{\text{n-pH}}:=(I_n,T\Ar T^{-1},T\Br,\Cr T^{-1},D)$, with form \eqref{eq:ABCD}. By defining  \cite{Mehrmann:2020}
\begin{equation}
    \mathbf S:=\matrixtwo{-T\Ar T^{-1}}{-T \Br}{\Cr T^{-1}}{D},
\end{equation}
one obtains the equivalent pH-form \eqref{eq:pH} by solving
\begin{equation}
    \matrixtwo{-J}{-G}{G^\top }{N}:= \dfrac{\mathbf S-\mathbf S^\top }{2} \text{ and }
    \matrixtwo{R}{P}{P^\top }{S}:= \dfrac{\mathbf S+\mathbf S^\top }{2}.
    \label{eq:pHmatrices}
\end{equation}

However, this approach suffers from two limitations. The first one stands in the assumption that the original model generating the data should be \emph{strictly} passive, and thus is cannot be applied to \emph{non-strictly} passive systems. The second one is more numerical: in practice Loewner pencil $(\sLL,\LL)$ of stable functions $\Htran$ may not be necessarily stable. Next, we propose two steps to overcome these limitations.

\subsection{Loewner framework with non-strict passivity and stability}
\label{sub:loe_passive2}

\subsubsection{Proposed modified algorithm}

Based on Algorithm 1 and 2 of \cite{Benner:2020}, we suggest the following Algorithm \ref{alg}, to identify pH model for non-strictly passive systems.

\begin{algorithm}
\caption{Data-driven normalized pH model construction of non-strictly passive system}\label{alg}
\begin{algorithmic}[1]
\Require $\{\lambda_j^0, \mathbf{r}_j^0, \mathbf{w}_j^0\}$, $\{\mu_i^0, \mathbf{l}_i^{0\top} , \mathbf{v}_i^{0\top} \}$, shift $D_s$ such that $D_s^\top+D_s\succ 0$, objective order $r$.
\Ensure $\Systemr_{\text{pH}}:=(M,Q,J,R,G,P,N,S)$ as in \eqref{eq:pH} and $\Systemr_{\text{pH}}$ ensuring interpolatory conditions.
\State Shift the data \eqref{eq:data} with $D_s$ as $\mathbf w_j\leftarrow \mathbf w_j^0+D_s$ and $\mathbf v_i\leftarrow \mathbf v_i^0+D_s$
\Comment{New step}
\State Construct the $r$-th order Loewner interpolant $\Systemr:=(-\LL,-\sLL,\mathbf V,\mathbf W,0_m)$ as in Section \ref {sub:loe} 
\State Compute the equivalent formulation $\Systemr:=(I_r,\Ar,\Br,\Cr,D_s)$ with transfer $\Htranr$
\State Compute projection $P_\infty(\Systemr)$ (or $P_\infty(\Htranr)$)
\Comment{New step}
\State Compute spectral zeros of $P_\infty(\Systemr)$ as in \eqref{eq:SZ} 
\State Set $\lambda_j\leftarrow \xi_j$, $\mathbf{r}_j\leftarrow \mathbf{x}_i$, $\mathbf{w}_j=\Htranr(\lambda_j)\mathbf r_j$ and \eqref{eq:data_strict_passive}
\State Construct $\LL$ and $\sLL$ as in \eqref{eq:loewner}
\State Construct $\sLL\leftarrow\sLL-\mathbf L D_s \mathbf R$, $\mathbf V\leftarrow \mathbf V-\mathbf L D_s$ and $\mathbf W\leftarrow \mathbf W-D_s\mathbf R$.
\State Compute Chloesky decomposition $\LL=T^\top T$
\State Construct $\Systemr_{\text{n-pH}}:=(I_n,T\Ar T^{-1},T\Br,\Cr T^{-1},D_s)$
\State Construct $\Systemr_{\text{pH}}:=(M,Q,J,R,G,P,N,S)$ using \eqref{eq:pHmatrices}
\State Set $S\leftarrow S-D_s$
\Comment{New step}
\end{algorithmic}
\end{algorithm}

The main modifications from \cite{Benner:2020} are listed hereafter. \emph{(i)} "Step 1", we suggest shifting the data with a positive scalar to translate the Nyquist response on the right hand side to ensure positive realness. This result in a data that are now strictly passive. \emph{(ii)} "Step 4", as the Loewner does not ensures stability, we suggest a projection of the rational ROM $\Htranr$ with realisation $\Systemr$ onto the  $\mathcal{RH}_\infty$ space, following \cite{Kohler:2014}. This leads to a stable $P_\infty(\Htranr)$ model. \emph{(iii)} "Step 12", based on the normalised realisation $\Systemr_{\text{n-Ph}}$, recover the original non-strictly passive model, by simply applying $S\leftarrow S-D_s$ after solving \eqref{eq:pHmatrices}, leading to the pH-ROM fitting the original data.

\subsubsection{Comments}

\paragraph{(i) and (iii)'s bullets} should the original model, and therefore associated data, be non strictly passive, but only passive. This typically occurs when no direct feed-through term exist. As a consequence, the resulting spectral zeros exhibit zeros on the imaginary axis. The first and last bullets address this point. One simply shifts the original problem to apply the strict-passive approach of \cite{Benner:2020}. 

\paragraph{(ii)'s bullet} the rational model obtained through the LF denoted $\Htranr$ may present unstable singularities. Therefore we suggest a \emph{post stabilisation} using the procedure presented in \cite{Kohler:2014}. This latter consists in projecting the rational model $\Htranr\in\mathcal{RL}_\infty$ onto its closest stable subset $\mathcal{RH}_\infty$, here using the $\mathcal H_\infty$-norm, leading to a stable model. Mathematically, and as exposed in details in \cite{Kohler:2014}, given a rational model $\Htranr\in\mathcal{RL}_\infty$ equipped with realisation $\Systemr$, one seeks $P_\infty(\Htranr) \in \mathcal{RH}_\infty$ such that,
\begin{equation}
P_\infty(\Htranr) = \arg \inf_{\mathbf G \in \mathcal{RH}_\infty}||\Htranr-\mathbf G||_{\mathcal{L}_\infty}.
\label{eq:stableApprox}
\end{equation}
Proof and procedure to obtain $P_\infty(\Htranr)$ are detailed in \cite{Kohler:2014}. The key steps consists in performing the stable and unstable part separation, then solving two reduced order Lyapunov equations. Applying this post-treatment to the Loewner-based approximate hopefully preserves the accuracy and interpolatory properties. Note that this step may also be addressed with \cite{Mehrmann:2020}. 

\paragraph{Interpolatory properties} At step 3, $\Htranr$ interpolates the \emph{shifted data} of step 1. At step 4, $\Htranr$ may not interpolate exactly these data due to the projection, but should likely do so if original system is stable. Then, at step 5 one should recover exactly $r$ strictly positive spectral zeros. Accordingly, the Cholesky decomposition at step 9 is possible as $\LL\succ 0$, thus at step 10 and 11, the system interpolates the \emph{spectral zeros data} of step 6, associated to the model $\Htranr$ constructed from the \emph{shifted data}. At step 10, the associated transfer function thus ensures $\Htranr_{\text{n-pH}}(\lambda_j^0)\mathbf r_j^0=\mathbf w^0_j+D_s$. Step 12 shifts back the model so that $\Htranr_{\text{pH}}$ tangentially interpolates the original data and is passive.

\section{Numerical use-case: the 2D wave equation}
\label{sec:num}

\subsection{Model description}

\subsubsection{Port-Hamiltonian formulation}
Let us consider the vertical deflection from equilibrium $w$ of a 2D membrane $\Omega \subset \mathbb{R}^2$. Denoting $\rho$ the mass density and $T$ the Young's modulus of the membrane, a positive-definite symmetric tensor, leads to the  damped wave equation given as \cite{KurZwa15} ($t \ge 0, \, x \in \Omega$)
$$
\rho(x) \frac{\partial^2}{\partial t^2} w(t,x) + \varepsilon(x) \frac{\partial}{\partial t} w(t,x) 
- {\rm div} \left( T(x) \cdot \grad \left( w(t,x) \right) \right) = 0
$$
where $\varepsilon$ is a positive damping parameter, together with Neumann boundary control $\left( T(x) \cdot \grad \left( w(t,x) \right) \right) \cdot \mathbf{n} = u_\partial(t,x)$, where $\mathbf{n}$ is the outward normal to $\Omega$. The Hamiltonian is the total mechanical energy, given as the sum of potential and kinetic energies
\begin{multline}
\mathcal{H}(t) := \frac{1}{2} \int_{\Omega} \left( \grad \left( w(t,x) \right) \right)^\top \cdot T(x) \cdot \grad \left( w(t,x) \right) {\rm d}x \\
+ \frac{1}{2} \int_{\Omega} \rho(x) \left( \frac{\partial}{\partial t} w(t,x) \right)^2 {\rm d}x.
\end{multline}
Taking the strain  $\aq := \grad \left( w \right)$ and the linear momentum $\alpha_p := \frac{\partial}{\partial t} w$ as energy variables, the Hamiltonian rewrites
\begin{multline}
\mathcal{H}(t) = \mathcal{H}(\aq (t,\cdot), \alpha_p(t,\cdot)) \\
= \frac{1}{2} \int_{\Omega} \left( \aq(t,x) \right)^\top \cdot T(x) \cdot \aq(t,x) {\rm d}x
+ \frac{1}{2} \int_{\Omega} \frac{\alpha_p^2(t,x)}{\rho(x)} {\rm d}x.
\end{multline}
The co-energy variables are by definition the variational derivatives of the Hamiltonian 
$\eq := \delta_{\aq} \mathcal{H} = T \cdot \aq$ the stress,  and $e_p := \delta_{\alpha_p} \mathcal{H} = \frac{1}{\rho}\alpha_p$,  the velocity. These equalities are the constitutive relations which close the dynamical system. Thanks to these variables, the wave equation writes as a port-Hamiltonian system
$$
\begin{pmatrix}
\frac{\partial}{\partial t} \aq \\
\frac{\partial}{\partial t} \alpha_p
\end{pmatrix} =
\begin{bmatrix}
0 & \grad \\
{\rm div} & - \varepsilon
\end{bmatrix}
\begin{pmatrix}
\eq \\
e_p
\end{pmatrix}, 
$$
In addition we denote inputs and outputs as $u_\partial = \eq \cdot \mathbf{n}\mid_{\partial \Omega}$ and $y_\partial = e_p\mid_{\partial \Omega}$. The power balance satisfied by the Hamiltonian is
\begin{equation}
    \label{eq-Hcontinu}
\frac{\rm d}{ {\rm d}t} \mathcal{H} = \langle u_\partial, y_\partial \rangle_{\partial \Omega} - \int_\Omega \varepsilon | e_p |^2
\leq 
\langle u_\partial, y_\partial \rangle_{\partial \Omega}\,,
\end{equation}
proving passivity. To get rid of the algebraic constraints induced by the constitutive relations, one rewrites the port-Hamiltonian system as
$$
\begin{bmatrix}
T^{-1} & 0 \\
0 & \rho
\end{bmatrix}
\begin{pmatrix}
\frac{\partial}{\partial t} \eq \\
\frac{\partial}{\partial t} e_p
\end{pmatrix} =
\begin{bmatrix}
0 & \grad \\
{\rm div} & -\varepsilon
\end{bmatrix}
\begin{pmatrix}
\eq \\
e_p
\end{pmatrix}, 
\; \left\lbrace
\begin{array}{rcl}
u_\partial &=& \eq \cdot \mathbf{n}, \\
y_\partial &=& e_p\mid_{\partial \Omega},
\end{array}\right.
$$
also known as the \emph{co-energy formulation}. 

\subsubsection{Structure-preserving discretization}

Let $\phiq$, $\varphi_p$ and $\psi$ be vector-valued, scalar-valued and boundary scalar-valued test functions respectively. The weak formulation reads
$$
\left\lbrace
\begin{array}{rcl}
\displaystyle \int_{\Omega} \phiq \cdot T^{-1} \cdot \frac{\partial}{\partial t} \eq 
&=& \displaystyle \int_{\Omega} \phiq \cdot \grad \left( e_p \right), \\
\displaystyle \int_{\Omega} \varphi_p \rho \frac{\partial}{\partial t} e_p 
&=& \displaystyle \int_{\Omega} \varphi_p {\rm div} \left( \eq \right)
- \int_{\Omega} \varphi_p \varepsilon e_p, \\
\displaystyle \int_{\partial \Omega} \psi y_\partial &=& \displaystyle \int_{\partial \Omega} \psi e_p.
\end{array}\right.
$$
The integration by parts of the second leads to ($u_\partial = \eq \cdot \mathbf{n}$) 
$$
\left\lbrace
\begin{array}{rcl}
\displaystyle \int_{\Omega} \phiq \cdot T^{-1} \cdot \frac{\partial}{\partial t} \eq 
&=& \displaystyle \int_{\Omega} \phiq \cdot \grad \left( e_p \right), \\
\displaystyle \int_{\Omega} \varphi_p \rho \frac{\partial}{\partial t} e_p 
&=& \displaystyle - \int_{\Omega} \grad \left( \varphi_p \right) \cdot \eq + \int_{\partial \Omega} \varphi_p u_\partial \\
&& \qquad \qquad - \int_{\Omega} \varphi_p \varepsilon e_p,\\
\displaystyle \int_{\partial \Omega} \psi y_\partial &=& \displaystyle \int_{\partial \Omega} \psi e_p.
\end{array}\right.
$$
Let $(\phiq^i)_{1 \le i \le N_q}$, $(\varphi_p^j)_{1 \le j \le N_p}$ and $(\psi^k)_{1 \le k \le N_\partial}$ be finite element families for $q$-type, $p$-type and boundary-type variables. Variables are approximated in their respective finite element family
$$
\eq^d(t,x) := \sum_{i=1}^{N_q} e_q^i(t) \phiq^i(x),
\qquad e_p^d(t,x) := \sum_{j=1}^{N_p} e_p^j(t) \varphi_p^j(x),
$$
$$
u_\partial^d(t,x) := \sum_{k=1}^{N_\partial} u_\partial^k(t) \psi^k(x),
\qquad y_\partial^d(t,x) := \sum_{k=1}^{N_\partial} y_\partial^k(t) \psi^k(x).
$$
Denoting $\underline{\star}$ the (time-varying) vector of coordinates of the discretisation $\star^d$ of $\star$ in its respective finite element family, the discrete system reads
$$
\small
\underset{M}{\underbrace{\begin{bmatrix}
M_q & 0 & 0 \\
0 & M_p & 0 \\
0 & 0 & M_\partial
\end{bmatrix} } }
\begin{pmatrix}
\frac{\rm d}{ {\rm d}t} \underline{e_q}(t) \\
\frac{\rm d}{ {\rm d}t} \underline{e_p}(t) \\
\ - \underline{y_\partial}(t)
\end{pmatrix} =
\underset{J-R}{\underbrace{\begin{bmatrix}
0 & G & 0 \\
\ -G^\top & - M_\varepsilon & B \\
0 & -B^\top & 0
\end{bmatrix} } }
\begin{pmatrix}
\underline{e_q}(t) \\
\underline{e_p}(t) \\
\underline{u_\partial}(t)
\end{pmatrix}
\normalsize
$$
where $(M_q)_{ij} := \int_{\Omega} \phiq^i \cdot T^{-1} \cdot \phiq^j$, $(M_p)_{ij} := \int_{\Omega} \varphi_p^i \rho \varphi_p^j$, $(M_\varepsilon)_{ij} := \int_{\Omega} \varphi_p^i \varepsilon \varphi_p^j$, $(M_\partial)_{ij} := \int_{\partial \Omega} \psi^i \psi^j$, and  $(B)_{jk} := \int_{\partial \Omega} \varphi_p^j\mid_{\partial \Omega}\,\psi^k$, $(G)_{ij} := \int_{\Omega} \phiq^i \cdot \grad \left( \varphi_p^j \right)$.
By definition, the discrete Hamiltonian is equal to the continuous Hamiltonian evaluated in the approximated variables. As we are working with the co-energy formulation, a first step is to restate the Hamiltonian in terms of co-energy variables, namely:
$$
\mathcal{H} = \frac{1}{2} \int_{\Omega} \eq \cdot T^{-1} \cdot \eq + \frac{1}{2} \int_\Omega \rho (e_p)^2.
$$
Then, the discrete Hamiltonian is defined as
$$
\mathcal{H}^d := \frac{1}{2} \int_{\Omega} \eq^d \cdot T^{-1} \cdot \eq^d + \frac{1}{2} \int_{\Omega} \rho (e_p^d)^2.
$$
After straightforward computations, it comes
$$
\mathcal{H}^d(t) = \frac{1}{2} \underline{e_q}(t)^\top M_q \underline{e_q}(t) + \frac{1}{2} \underline{e_p}(t)^\top M_p \underline{e_p}(t),
$$
and the \emph{discrete} power balance follows
\begin{eqnarray*}
\frac{\rm d}{ {\rm d}t} \mathcal{H}^d(t) &=& \underline{u_\partial}(t)^\top M_\partial \underline{y_\partial}(t) - \underline{e_p}(t)^\top M_\varepsilon \underline{e_p}(t) \\
&\leq& 
\underline{u_\partial}(t)^\top M_\partial \underline{y_\partial}(t)
\,,
\end{eqnarray*}
mimicking \eqref{eq-Hcontinu} exactly at the discrete level. The pH-FOM is thus given by the a realization $\System_{\textbf{pH}}$ in the form \eqref{eq:pH}. The  convergence of the numerical method is assessed in \cite{HaineIJNAM2023}, where the optimal selection of families of Finite Elements is proved. The objective is to use these matrices and the associated transfer function $\Htran$ to generate \emph{data}, through the dedicated SCRIMP simulator\footnote{See {\tt https://g-haine.github.io/scrimp/}}, presented in \cite{SCRIMP2021}. These data shall serve the construction of a pH-ROM, as explained in \S\ref{sec:rom}.

\subsection{pH-ROM identification}

The  considered model $\Htran$ is a 2D wave equation, on an L-shaped domain $\Omega$. The resulting discretized pH model is equipped with a realisation $\SystemPh$ in pH-form as in \eqref{eq:pH}. It has the following characteristics: $n=63,409$ and $m=604$. For illustration, we will now consider to sub-cases. First, \emph{(i)} the SISO case: $m=1$, where we consider the first input / output pair only and second, \emph{(ii)} the MIMO case: $m=3$, where we consider the following input / output index pairs $\{1,2,496\}$ only. Notice that as the system is collocated, transfers one and two are highly close while the third one is far away and will thus be very different. The model is stable, poorly damped, but not strictly passive, thus spectral zeros on $\imath \mathbb R$ should unlikely occur, resulting in a non verified Assumption~\ref{ass:strict_passivity}. In what follows, we denote as: 
\begin{itemize}
    \item \emph{Data}, the original system presented in \S \ref{sec:num}-A-B sampled along $\imath \omega_l$, where $\omega_l$ are $l=1,\dots,300$ logarithmically spaced values between $10^{-1}$ and $10^{3.5}$ rad/s.
    \item \emph{Loewner}, the dynamical model obtained with the standard LF, in the form \eqref{eq:EABC}.
    \item \emph{pH-Loewner}, the dynamical model obtained with the Algorithm \ref{alg} (input data being closed conjugated from \emph{data} and $D_s=1$), in the form \eqref{eq:pH}.
\end{itemize} 
In each case, we also use the denomination \emph{shifted} to point the data or model shifted and with \emph{post-stability enforcement}, to ensure strict dissipativity.

\begin{figure}
    \centering
    \includegraphics[width=\columnwidth]{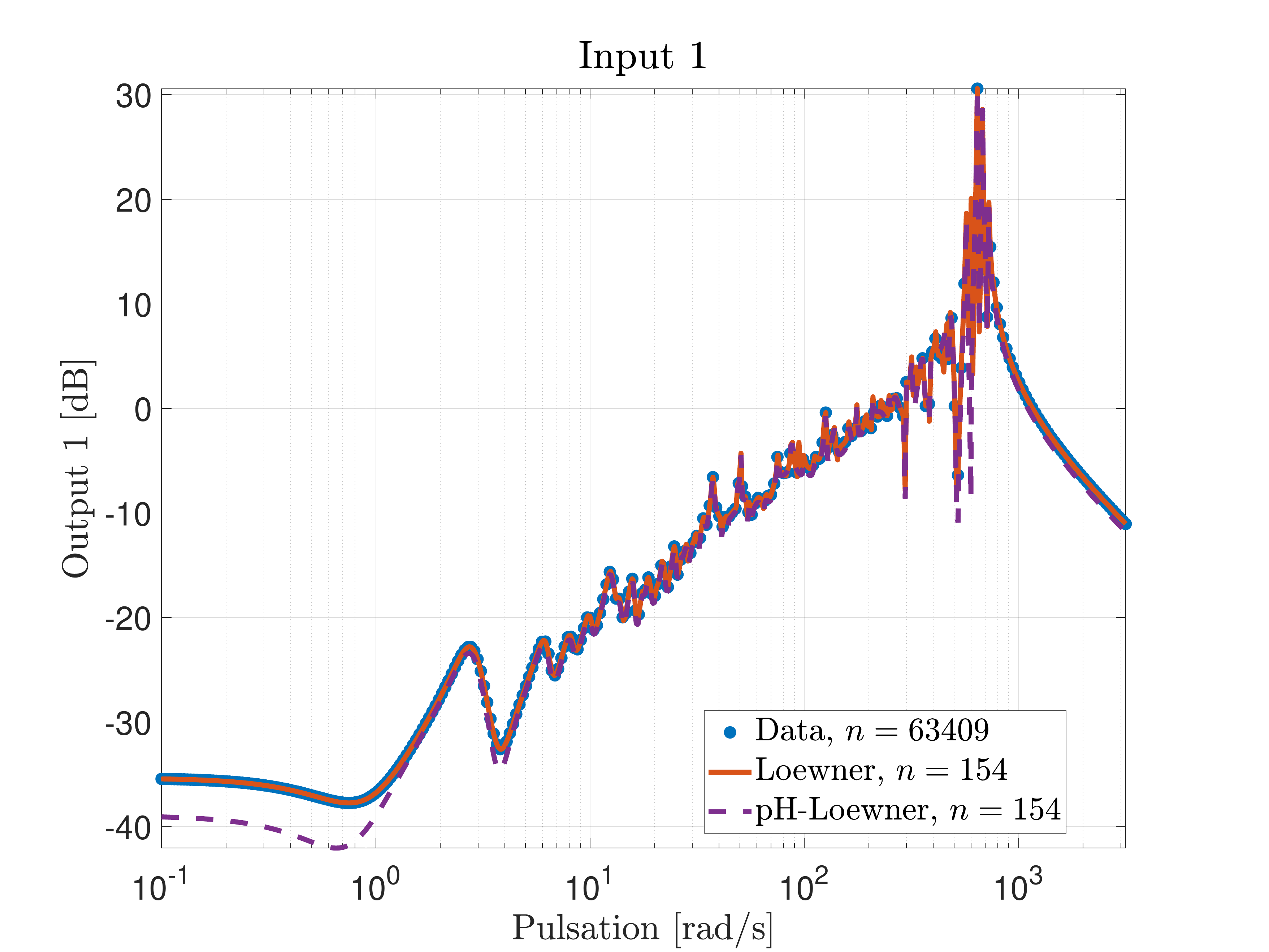}
    \caption{Frequency response of the original data, the Loewner and the pH-Loewner models.}
    \label{fig:fr_siso}
\end{figure}

\subsubsection{SISO case: process illustration}

first, Figure \ref{fig:fr_siso} presents the frequency response of the original \emph{data}, compared to the (non-dissipative and non-stable) \emph{Loewner} model and the \emph{pH-Loewner} one, illustrating the nice restitution of the frequency response in both cases. Here, the  \emph{pH-Loewner} is passive and embeds the expected pH-structure. 

\begin{figure}
    \centering
    \includegraphics[width=\columnwidth]{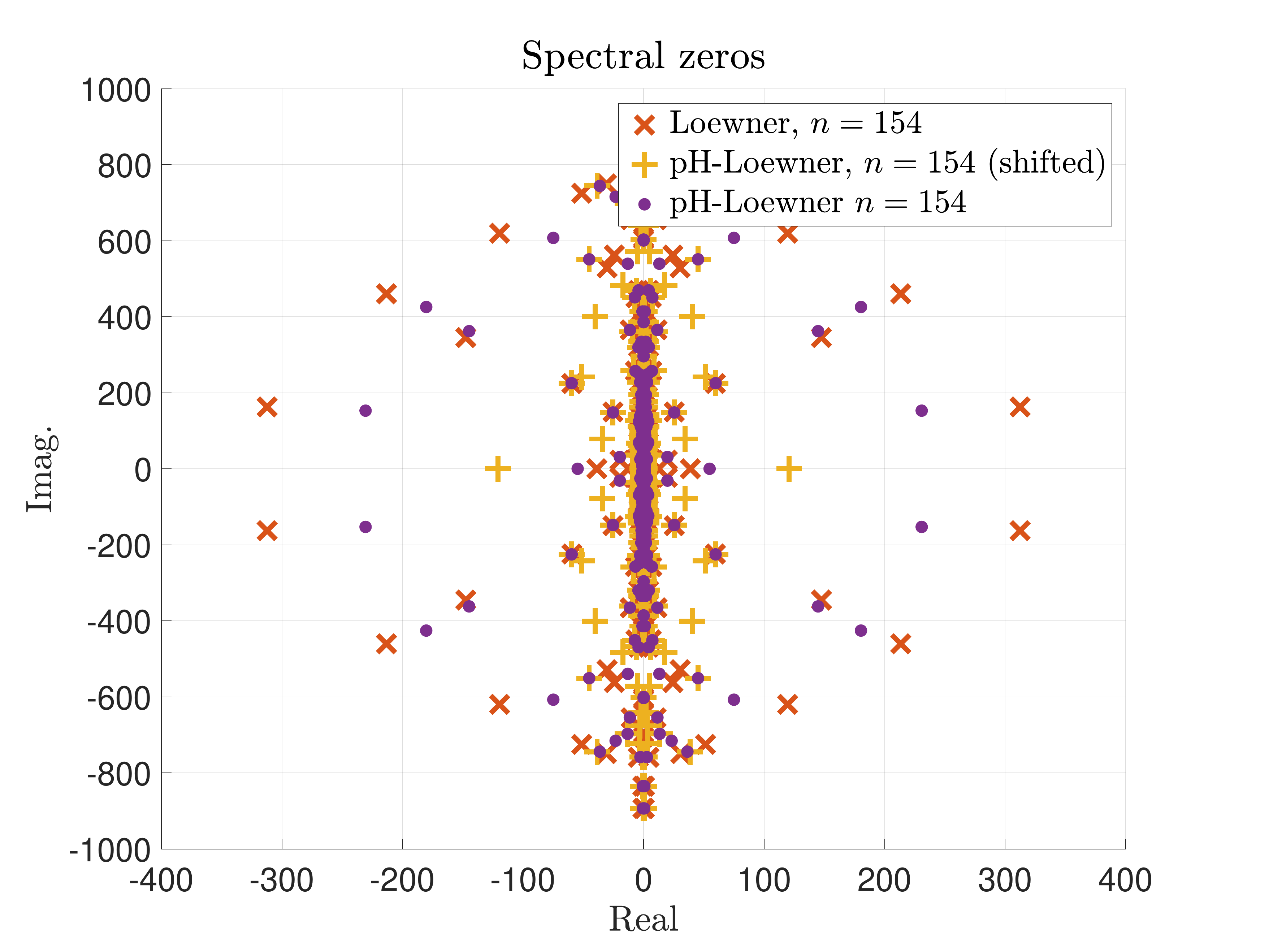}
    \caption{Spectral zeros response of the Loewner, the Loewner (applied on the shifted data) and the pH-Loewner models.}
    \label{fig:sz_siso}
\end{figure}

\begin{figure}
    \centering
    \includegraphics[width=\columnwidth]{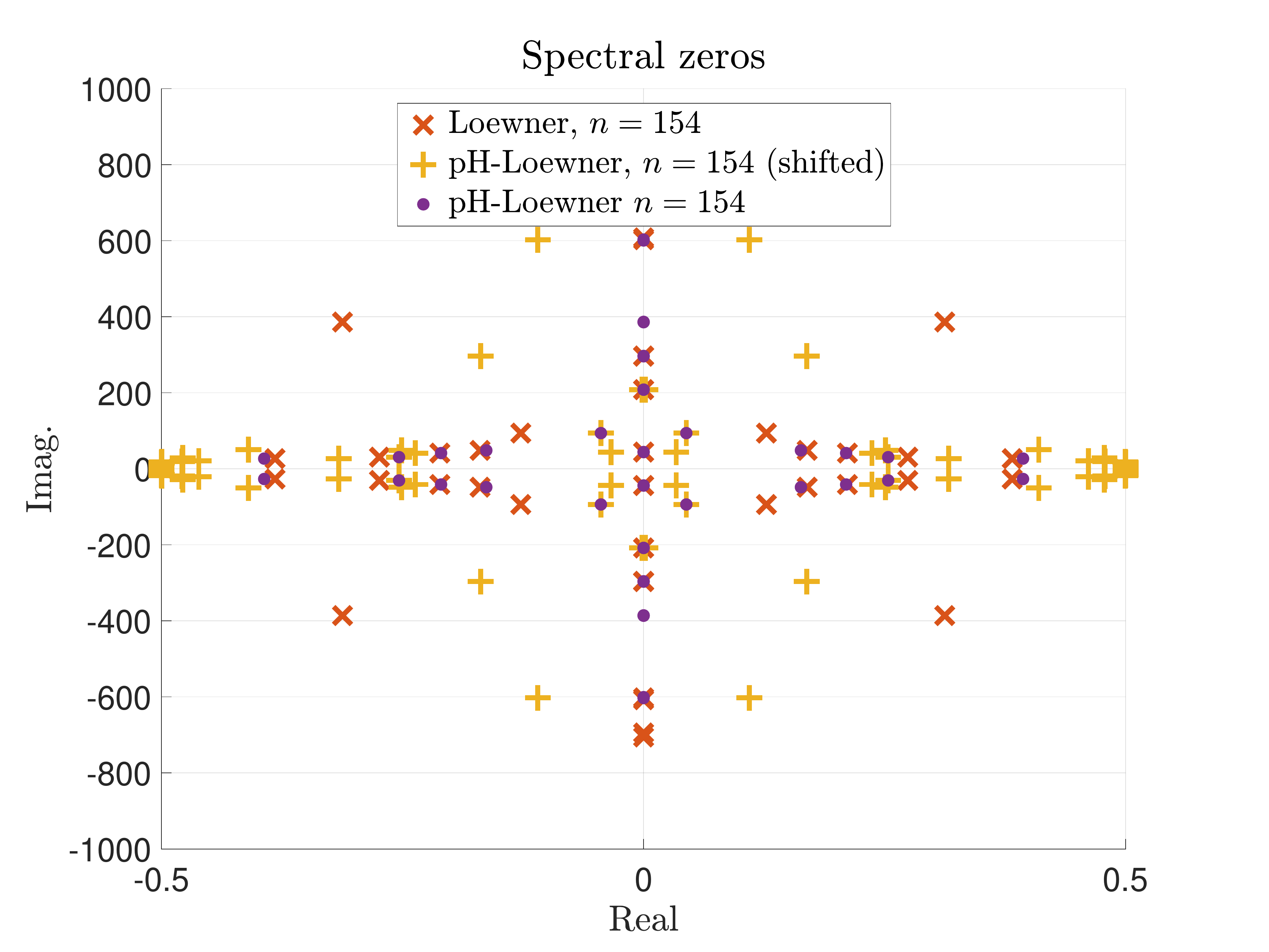}
    \caption{Spectral zeros response of the Loewner, the Loewner (applied on the shifted data) and the pH-Loewner models.}
    \label{fig:sz_siso_zoom}
\end{figure}

Figure \ref{fig:sz_siso} (\ref{fig:sz_siso_zoom}) show the (zoomed) spectral zeros. The \emph{Loewner} model has many zeros along or close to the imaginary axis ($\boldsymbol \times$). Indeed,  between the real band $[-1,1]\times 10^{-10}$, we count 4 zeros and between $[-1,1]\times 10^{-9}$, 12. This is an issue for selecting the positive interpolation points. This problem is solved by the proposed algorithm modification, thanks to both \emph{post-stability enforcement} and \emph{data-shift}. Indeed, the \emph{pH-Loewner (shifted)} shows zeros far from this limit ($\boldsymbol +$). Then, after applying the shift-back, one recovers the sought \emph{pH-Loewner} model ($\bullet$), where spectral zeros are back on the imaginary axis. Note that without the proposed algorithm adjustments, no solution can be found as the $\LL$ matrix is not positive definite and Cholesky decomposition is impossible. A reduced order model can be obtained if at step 2 of Algorithm \ref{alg} one selects a $r<154$. In the presented figures,  the order selection was performed by the rank revealing decomposition of the Loewner matrices.

\begin{figure}
    \centering
    \includegraphics[width=\columnwidth]{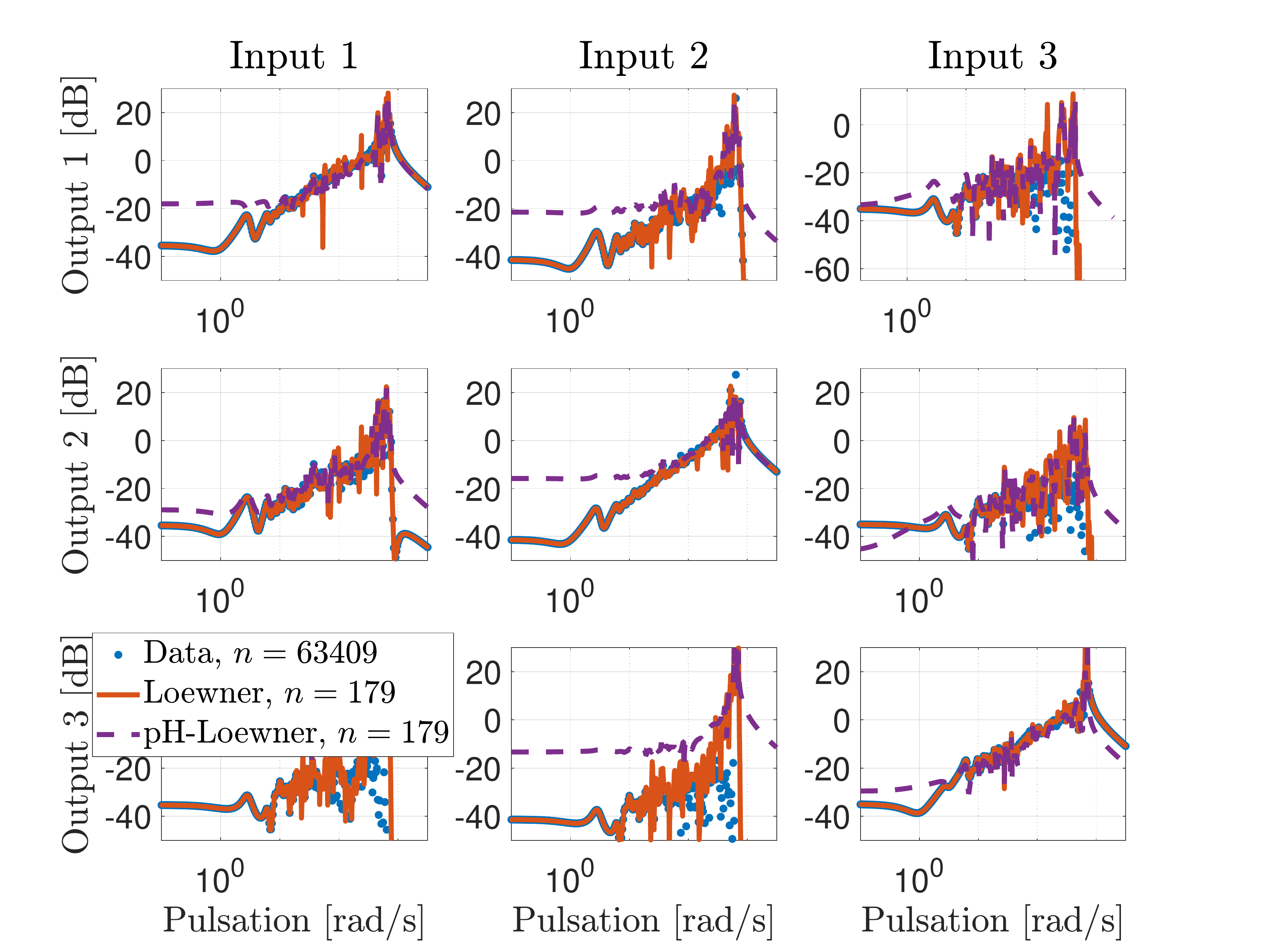}
    \caption{Frequency response of the original data, the Loewner and the pH-Loewner models.}
    \label{fig:fr_mimo_rom}
\end{figure}

\subsubsection{MIMO case}

as mentioned, one important challenge in this application in addition to the complexity of the dynamics of the wave equation, is its large number of inputs and outputs ($m=604$). So far, applying the above process to this large MIMO system led to non fully satisfactory results. However, up to $m=10$, reasonably good approximation have been observed. In Figure \ref{fig:fr_mimo_rom}, we illustrate the frequency magnitude response for $m=3$, where the first two  inputs / outputs are spatially close, whereas the third one is spatially far from the others. 

Figure \ref{fig:fr_mimo_rom} illustrates the fact that diagonal elements (more energetic since the model is collocated) are well reproduced. Regarding the anti-diagonal ones, a good restitution is observed on channels (1,2) and (2,1) but not on (1,3), (2,3), (3,1) and (3,2). Indeed, these transfer are less energetic than the other and thus not well approximated. This is a point for future investigations, \eg with a specific treatment of the tangential directions in \eqref{eq:data}.

\section{Conclusions \& perspectives}
\label{sec:conclusion}

We have shown promising numerical  results of a \emph{data-driven} reduction technique applied to a 2D wave PDE, modelled as a pHs: it is obtained using the LF, and more specifically a modification of \cite{Benner:2020} using the frequency-responses generated by the matrices provided by the structure-preserving PFEM. The main modifications to \cite{Benner:2020} are  \emph{(i)} the data-shift to handle non-strictly passive models and \emph{(ii)} the post-stability enforcement, to cope with numerical issues often encountered when applying the LF. These two steps where essential to achieve the presented results. Indeed, what is successful is the number of states that can be drastically reduced (from $n=63,409$ to $n=179$). However, the collocated input-output pairs have been tried on a SISO case, or on a MIMO case of small dimension ($m=3,\dots,10$). So far, the MIMO version remains not fully satisfactory and will be of specific attention in future researches. 
Further investigations will also consider handling a larger number of inputs-outputs and different real world applications\footnote{See e.g. \url{https://algopaul.github.io/PortHamiltonianBenchmarkSystems.jl/}}, such as Maxwell equations in 2D or even in 3D, see \cite{MonteghettiMTNS2022}.

\bibliographystyle{ieeetr}
\bibliography{biblio}
\end{document}